\documentclass[lettersize,journal]{IEEEtran}
\usepackage{amsmath,amsfonts}
\usepackage{algorithmic}
\usepackage{algorithm}
\usepackage{array}
\usepackage{textcomp}
\usepackage{stfloats}
\usepackage{url}
\usepackage{verbatim}
\usepackage{graphicx}
\usepackage{cite}
\usepackage{subcaption}
\usepackage{amssymb}
\usepackage{tikz}
\usetikzlibrary{fit,positioning}
\usepackage{pgfplots}
\pgfplotsset{compat=1.18}
\usepackage{xcolor}
\usepackage{xfrac}

\usepackage[acronym,nomain]{glossaries}
\newacronym{pai}{physical AI}{physical artificial intelligence}
\newacronym{ber}{BER}{bit error rate}
\newacronym{eier}{EIER}{event-inference error ratio}
\newacronym{twi}{TWI}{temporal window of integration}
\newacronym{eir}{EIR}{event-inference reliability}
\newacronym{snr}{SNR}{signal to noise ratio}
\newacronym{sinr}{SINR}{signal to interference plus noise ratio}
\newacronym{cdf}{CDF}{cumulative distribution function}
\newacronym{bs}{BS}{base-station}
\newacronym{aoi}{AoI}{age-of-information}
\newacronym{mi}{MI}{mutual-information}
\newacronym{map}{MAP}{maximum a posteriori}
\newacronym{sr}{SR}{scheduling request}
\newacronym{pt}{PT}{packet transmission}
\newacronym{pmf}{PMF}{probability mass function}
\newacronym{mse}{MSE}{mean-square error}
\newacronym{lmmse}{LMMSE}{linear minimum mean-square error}
\newacronym{mira}{MIRA}{maximum information-rate allocation}
\newacronym{rb}{RB}{resource block}
\newacronym{mcs}{MCS}{modulation and coding scheme}
\newacronym{mia}{MIA}{maximum information allocation}
\newacronym{se}{SE}{spectral efficiency}
\newacronym{tti}{TTI}{transmission time interval}

\begin{document}

\title{Wireless Evidence Acquisition for Multimodal Inference \vspace{-0.1cm}\\Constrained by Temporal Windows}

\author{
Alessandro Compagnoni,~\IEEEmembership{Student Member,~IEEE},
Anup Mishra,~\IEEEmembership{Member,~IEEE},
Carla Fabiana Chiasserini,~\IEEEmembership{Fellow,~IEEE},
Elad Michael Schiller,
and Petar Popovski,~\IEEEmembership{Fellow,~IEEE}
\vspace{-0.6cm}
\thanks{A. Compagnoni and C. F. Chiasserini are with the Department of Electronics and Telecommunications, Politecnico di Torino, 10129 Torino, Italy (e-mail: alessandro.compagnoni@polito.it; carla.chiasserini@polito.it).}%
\thanks{A. Mishra and P. Popovski are with the Department of Electronic Systems, Aalborg University,
Aalborg, Denmark (e-mail: anmi@es.aau.dk; petarp@es.aau.dk).}%
\thanks{C. F. Chiasserini and E. M. Schiller are also with
the Department of Computer Science and Engineering, Chalmers University of Technology and the University of
Gothenburg, Gothenburg, Sweden (e-mail: elad@chalmers.se).}%
}



\maketitle

\begin{abstract}
Multimodal inference in safety-critical applications, such as autonomous driving and robot navigation, requires heterogeneous sensor observations to reach an edge server within a task-prescribed temporal window. Referenced to the event or state being inferred, this window ends at the latest time an observation remains useful for the current decision. Since sensor readiness times, data volumes, wireless channel conditions, and task relevance vary across sensors, maximising network throughput does not necessarily minimise the fused prediction error when the window closes. We cast multimodal uplink scheduling as sequential wireless evidence acquisition and, for a \gls{lmmse} fusion model, derive a conditional evidence gain metric from the reduction in residual-error volume. Defined through second-order statistics, this metric applies beyond Gaussian models and coincides with conditional mutual information when the target and prediction errors are jointly Gaussian. We then develop {MIRA}, a greedy task-driven \emph{maximum information-rate allocation} policy that combines conditional evidence gain with each sensor's channel state information and remaining data volume, while updating sensor relevance as evidence is acquired. Experiments on synthetic classification, human activity recognition, and vehicle-trajectory regression show that MIRA outperforms both relevance-only and channel-only scheduling. Relative to the former, MIRA improves classification accuracy by up to $70\%$ and reduces regression \gls{mse} by up to $5.5\%$. Relative to the latter, it requires $58\%$ less acquisition time to attain an $80\%$ target accuracy, while achieving up to $95\%$ higher accuracy and $9\%$ lower regression \gls{mse}. These gains are achieved without maximising received-data volume.
\end{abstract}


\section{Introduction}
Wireless-enabled \gls{pai} systems adopt heterogeneous sensors to infer and act upon the
physical world~\cite{Wu2026PhysicalAI,Ahuja_MML}. At the decision
node, visual, acoustic, inertial, or locally extracted features
collectively constitute the \emph{evidence} supporting a
prediction. Task-oriented communication already recognises that
wireless design should be evaluated through downstream utility
rather than delivered bits alone~\cite{strinati2021beyond,Shao}.
In multimodal inference, however, wireless decisions also determine
the composition of the evidence itself by controlling which
observations become available and when.
The usefulness of this evidence is inherently time-dependent. 

Value-of-information scheduling captures this principle by relating the value of an update to its timeliness and inferential consequences~\cite{chiariotti2022voi,shisher2023learning}. 
For event-driven multisensory tasks, the \gls{twi} framework, inspired by human multisensory perception, anchors observation timing to the underlying physical event and determines when distributed observations can be integrated for a common  decision~\cite{mishra2025temporal}. It distinguishes the task-dependent usefulness of each observation from cross-modal simultaneity and causal consistency, while \gls{eir} characterises how much uncertainty the admitted evidence resolves~\cite{mishra2026eir}. Together, \gls{twi} and \gls{eir} provide temporal and inferential foundations for evaluating multisensory evidence. 
Operationalising these foundations over a wireless network requires deciding which candidate observations should be delivered before their usefulness expires when the available radio resources are insufficient to acquire them all. Information-based sensor-selection methods address a related problem by identifying informative sources while accounting for redundancy, complementarity, and acquisition costs~\cite{krause2008near,ballotta2019computation}. For event-driven wireless inference, however, evidence acquisition unfolds as a sequential service process: sensors become ready at different times, generate data objects of different sizes, and experience time-varying channels. Each allocation therefore reshapes the remaining acquisition opportunity and, upon completion of an observation, updates the conditional value of the evidence still awaiting service.
Within the \gls{twi} framework, the operational coupling between inferential value and wireless acquisition remains largely unexplored. The work in~\cite{lin2026joint} advances one direction by combining usefulness-based window selection with slot-level wireless control to maintain an inferential state over time. We pursue the complementary direction of constructing evidence for a common prediction within a prescribed \gls{twi}. At each slot, the scheduler must balance the additional prediction value of a candidate observation, given the evidence already acquired, against the radio opportunity required to obtain it.

\vspace{+0.015cm}
\noindent
{\bf Contributions.} (i) We cast multimodal uplink scheduling within a prescribed \gls{twi} as sequential wireless evidence acquisition and formulate the corresponding allocation problem for heterogeneous sensor readiness times, data volumes, and channel conditions, aiming to minimise the fused prediction \gls{mse} at the end of the \gls{twi}. (ii) For an \gls{lmmse} fusion model, we derive an information-based metric from the uncertainty volume reduction, which coincides with conditional mutual information under joint Gaussianity. (iii) We develop \gls{mira}, a greedy policy that translates this gain into a wireless acquisition priority based on \gls{se} and remaining data volume, while updating modality relevance as new evidence is acquired. (iv) We evaluate \gls{mira} on synthetic multimodal classification, real-world human activity recognition, and vehicle-trajectory regression. Results show that \gls{mira} outperforms traditional sum-rate-maximising scheduling, particularly when modality relevance and channel quality are misaligned, with gains exceeding 95\%, while also demonstrating that acquiring more data does not necessarily improve inference performance. 

\vspace{+0.045cm}
\noindent
{\bf Notation.} $\boldsymbol{X}\in\mathbb{R}^N$ denotes a \emph{column} random vector taking values in $\mathbb{R}^N$, with $i$-th element $X[i]$. $\mathbf{I}_N$ is the $N{\times} N$ identity matrix. $||\boldsymbol{X}||$ denotes $\boldsymbol{X}$'s Euclidean norm. $\boldsymbol{A}^{\mathsf{T}}$, $\boldsymbol{A}^{-1}$, and $\boldsymbol{A}^{+}$ denote (resp.) the transpose, inverse, and pseudoinverse of the matrix $\boldsymbol{A}$. $\lceil\,\cdot\,\rceil$ and $\lfloor\,\cdot\,\rfloor$ denote (resp.) rounding up and down to the nearest integer. $\boldsymbol{\Sigma}_{\boldsymbol{X}\boldsymbol{Y}}=\mathbb{E}[(\boldsymbol{X}-\mathbb{E}[\boldsymbol{X}])(\boldsymbol{Y}-\mathbb{E}[\boldsymbol{Y}])^{\mathsf{T}}]$ denotes the cross-covariance of real-valued random vectors $\boldsymbol{X},\boldsymbol{Y}$, with $\mathbb{E}[\,\cdot\,]$ being the expectation operator. 
\begin{figure}[t]
    \centering
    \includegraphics[
        width=1\linewidth,
        trim={0.85cm 0cm 0cm 0cm},
        clip
    ]{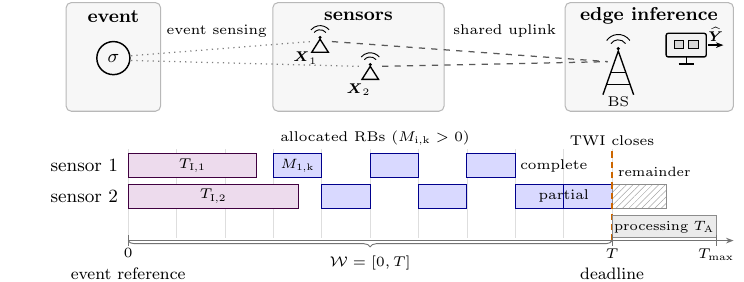}
    \caption{{\small Wireless evidence acquisition within a prescribed
    \gls{twi}. Some of sensor 2’s data arrive after $T$ and cannot be used for prediction.\vspace{-0.5cm}}}
    \label{fig:diagram}
\end{figure}

\section{System Model and Problem Formulation}\label{sec:system_model}
We consider a physical event $\sigma$ and an associated target random vector $\boldsymbol{Y}\in\mathbb{R}^{C}$ to be predicted. The task prescribes a \gls{twi} with usefulness interval $\mathcal{W}=[t_e,t_e+T]$, where $t_e=0$ is the event reference (set to $0$ without loss of generality~\cite{mishra2025temporal,mishra2026eir}) and $T$ is the task-prescribed usefulness horizon, whose endpoint serves as the evidence-acquisition deadline for the current prediction. We assume that the candidate observations satisfy the event-association, cross-modal simultaneity, and causal-consistency conditions of the \gls{twi} framework. Thus, an observation is admitted only if it is delivered by the evidence-acquisition deadline $T$, at which the \gls{twi} closes. If inference requires a fixed duration $T_A$, the application deadline is $T_{\max}=T+T_A$, when the final prediction is produced. The event is observed by a set of heterogeneous sensors $\mathcal{S}{=}\{s_1,\ldots,s_S\}$. Sensor $s_i$ requires a sensor-specific modality-dependent integration time $T_{I,i}$ to obtain its observation and produce an $N_i$-dimensional data object $\boldsymbol{X}_i$ (a random vector) of size $b_i^{\star}$ bits, taking values in an alphabet $\mathcal{X}_i$. Upon completion of the measurement at $t=T_{I,i}$, $\boldsymbol{X}_i$ becomes
available for uplink transmission to a \gls{bs} co-located with the edge inference node. Because the sensors share finite radio resources, the allocation policy determines how their data objects are served within $\mathcal{W}$ and, consequently, the sensor-derived evidence available when the window closes; see Fig.~\ref{fig:diagram}. The inference node processes the acquired observations using sensor-specific prediction models and combines their outputs to estimate $\boldsymbol{Y}$.
To characterise how the acquired evidence supports prediction, the inference node applies a prediction model $r_i(\cdot)$ associated with the modality observed by sensor $s_i$, producing
$\widehat{\boldsymbol{Y}}_i
=r_i(\boldsymbol{X}_i)
=\boldsymbol{Y}+\boldsymbol{\Upsilon}_i$, where $\boldsymbol{\Upsilon}_i\in\mathbb{R}^{C}$ is the corresponding prediction error. Sensors covering the same modality may employ the same prediction model. We define $\mathcal{A}(t)
\triangleq
\{
s_i\in\mathcal{S}:
\boldsymbol{X}_i\text{ has been fully received by time }t
\}$ as the acquired sensor set at time $t$. Hence sensor $s_i$ joins this set only after its entire data object $\boldsymbol{X}_i$ has been received. Accordingly, $\mathcal{A}(T)$ identifies the sensor-derived evidence constructed by the allocation policy when the \gls{twi} closes; hence any data not delivered by $T$ are excluded from the current prediction.

For any nonempty set
$\mathcal{A}\subseteq\mathcal{S}$
of sensors whose complete observations are available, we stack the
corresponding predictions as $\boldsymbol{P}_{\mathcal{A}}
    \triangleq [\widehat{\boldsymbol{Y}}_{i}^{\mathsf{T}}
    ]^{\mathsf{T}}_{i:s_i\in \mathcal{A}}
    \in\mathbb{R}^{|\mathcal{A}|C}$. 
    
The means and cross-covariances of $\boldsymbol{Y}$ and the
complete-observation predictions, including the statistical
dependence between their errors, are estimated offline from
calibration data and assumed to be available at the inference node.
The corresponding \gls{lmmse} estimate is
\begin{equation}
    \widehat{\boldsymbol{Y}}_{\mathcal{A}}
    =
    \mathbb{E}[\boldsymbol{Y}]
    +
    \boldsymbol{\Sigma}_{
        \boldsymbol{Y}\boldsymbol{P}_{\mathcal{A}}
    }
    \boldsymbol{\Sigma}_{
        \boldsymbol{P}_{\mathcal{A}}
        \boldsymbol{P}_{\mathcal{A}}
    }^{+}
    \left(
        \boldsymbol{P}_{\mathcal{A}}
        -
        \mathbb{E}[\boldsymbol{P}_{\mathcal{A}}]
    \right).
    \label{eq:lmmse_fusion}
\end{equation}
For $\mathcal{A}=\emptyset$, we set
$\widehat{\boldsymbol{Y}}_{\emptyset}
=\mathbb{E}[\boldsymbol{Y}]$.
The model applies to both regression and classification. For
the latter, $\boldsymbol{Y}$ is one-hot encoded, and each
$\widehat{\boldsymbol{Y}}_i$ is a softmax-output vector.

Since sensors differ in relevance, the composition of $\mathcal{A}$ directly affects accuracy: under limited bandwidth, it is crucial to prioritise informative data, yet the most informative sensors may also suffer poor channels, making resource allocation to them potentially inefficient. These aspects are addressed jointly by the proposed MIRA method in Sec.~\ref{sec:method}.

\vspace{+0.045cm}
\noindent
{\bf Slot-Based Wireless Evidence Acquisition.} The wireless evidence-acquisition interval
$\mathcal{W}{=}[0,T]$ is divided into \glspl{tti} of duration $T_s$, such that the $k$-th \gls{tti} starts at time $t {=} k T_s$, with $k {=} 0,1,\ldots,K{-}1$ and $K{=}\lfloor T/T_s\rfloor$.
We adopt slot-based scheduling and assume that the resource allocation remains fixed during each \gls{tti}.
Sensor $s_i$ becomes eligible
for transmission starting from slot $\kappa_i
    \triangleq
    \left\lceil{T_{I,i}}/{T_s}\right\rceil$,
after which its data object of size $b_i^\star$ bits is placed in
the transmission buffer. Let $b_{i,k}$ denote its remaining data
volume at the beginning of slot $k$, and define
$\mathcal{D}_k
    \triangleq
    \{
        s_i{\in}\mathcal{S}:
        k{\geq}\kappa_i,
        b_{i,k}{>}0
    \}$ as the set of sensors that are ready and have data awaiting
transmission. At every slot $k$, the \gls{bs} allocates $M$ \glspl{rb} among
the sensors in $\mathcal{D}_k$. Let $M_{i,k}\in\mathbb{Z}_{\geq0}$
be the number of \glspl{rb} assigned to sensor $s_i$.
Under independent block, flat Rayleigh fading, the received \gls{snr} for sensor $s_i$ in slot $k$ is $\rho_{i,k}=\gamma_i g_{i,k}$, where $g_{i,k}{\sim}\operatorname{Exp}(1)$ and $\gamma_i$ is its average \gls{snr}. Link adaptation maps $\rho_{i,k}$ to a \gls{mcs} with \gls{se}
$\xi_{i,k}$ (bps/Hz). In practical 5G NR systems, the uplink \gls{mcs} is selected to achieve a target first-transmission transport block error rate, typically around $10\%$ \cite{TS38214}. In this paper, \gls{se} values are derived according to \cite{markova2024performance}, which provides a representative mapping among SNR, channel quality indicators (CQIs), and \gls{se} in 5G NR systems.

With \gls{rb} bandwidth $B$, the resulting transmission rate for sensor $s_i$ is $R_{i,k}
    =
    \xi_{i,k}M_{i,k}B$ (bps).

A sensor $s_i$ joins the completely acquired set $\mathcal{A}(t)$ when
its buffer reaches zero. To accommodate partial reception, we assume that the elements of
$\boldsymbol{X}_i$ are encoded and transmitted sequentially, with
each element requiring roughly $b_i^\star/N_i$ bits. Let
$\ell_i{\leq} N_i$ denote the number of elements received from sensor $s_i$ by the
end of $\mathcal{W}$. If $\ell_i{=}N_i$, the complete observation
is available. Otherwise, the missing elements
$(X_i[\ell_i+1],\ldots,X_i[N_i])$ are replaced with their
training-set means, yielding an imputed sequence denoted by
$\overline{\boldsymbol{X}}_i$. Using the evidence available at $T$, predictions obtained from
the complete and imputed observations are supplied to the fusion
rule in~\eqref{eq:lmmse_fusion}. During scheduling,
$\mathcal{A}(t)$ contains only fully received observations and is
used to update the conditional evidence gains defined in Sec.~\ref{sec:method}, whereas
$\{\ell_i\}_{i=1}^{S}$ characterises the partially received
evidence used in the final fusion. Data arriving after $T$ are
excluded from the current prediction.

\vspace{+0.045cm}
\noindent
{\bf Sequential Evidence-Acquisition Problem.} At the beginning of slot $k$, the \gls{bs} observes the current
allocation state $\Omega_k
    {\triangleq}
    \left\{
        \mathcal{D}_k,\,
        \mathcal{A}_k,\,
        \boldsymbol{b}_k,\,
        \boldsymbol{\xi}_k
    \right\}$, where $\mathcal{A}_k{\triangleq}\mathcal{A}(kT_s)$,
$\boldsymbol{b}_k{=}[b_{1,k},\ldots,b_{S,k}]^{\mathsf{T}}$, and
$\boldsymbol{\xi}_k$ contains the spectral efficiencies observed
in slot $k$. We denote the state history available by slot $k$ as
$\Omega_{0:k}
{=}\{\Omega_0,\ldots,\Omega_k\}$.
Let $\Psi_k$ be the slot-$k$ allocation rule, which maps
$\Omega_{0:k}$ to a specific \gls{rb}-allocation vector $[M_{i,k}]^{\mathsf{T}}_{i:s_i \in \mathcal{D}_k}$. The resulting causal
allocation policy over the prescribed \gls{twi} is given by
$\Psi=\{\Psi_0,\ldots,\Psi_{K-1}\}$. Denoting by $\mathcal{P}$ the set of feasible policies and by $\widehat{\boldsymbol{Y}}_{\Psi}$ the estimate produced at $T_{\max}$ from the evidence acquired by $T$ using~\eqref{eq:lmmse_fusion}, the optimal resource-allocation problem pursued in this work is $\Psi^{\star}
    =
    \arg\min_{\Psi\in\mathcal{P}}
    \mathbb{E}_{\Psi}[
        ||
            \boldsymbol{Y}
            -
            \widehat{\boldsymbol{Y}}_{\Psi}
        ||^2
    ]$, where the expectation is over the sensor observations, readiness
times, and wireless-channel realisations under policy $\Psi$. This problem is sequential as each
slot-level allocation changes the evidence that can be acquired before the
\gls{twi} closes. Solving it requires optimising over causal
allocation rules for the possible evolutions of sensor readiness,
wireless channels, and acquired evidence. This motivates the greedy policy developed in
Sec.~\ref{sec:method}.
\section{The MIRA Policy}
\label{sec:method}
To construct a tractable greedy policy for the terminal prediction
objective, we quantify the value of completing each candidate
observation through its conditional reduction of residual-error
volume. At slot $k$, let $\mathcal{A}_k$ denote the sensors whose complete
observations have already been acquired. For a candidate sensor
$s_i\in\mathcal{S}\setminus\mathcal{A}_k$, we first quantify the
inferential benefit that would result from completing its data
object, given the evidence already available.
For an acquired set $\mathcal{A}$, let
$\boldsymbol{E}_{\mathcal{A}}
    \triangleq
    \boldsymbol{Y}
    -
    \widehat{\boldsymbol{Y}}_{\mathcal{A}}$ denote the residual prediction error associated with complete
observations from $\mathcal{A}$. Its covariance is
$\boldsymbol{\Sigma}_{
        \boldsymbol{E}_{\mathcal{A}}
        \boldsymbol{E}_{\mathcal{A}}
    }
    {=}
    \boldsymbol{\Sigma}_{\boldsymbol{Y}\boldsymbol{Y}}
    -
    \boldsymbol{\Sigma}_{
        \boldsymbol{Y}\boldsymbol{P}_{\mathcal{A}}
    }
    \boldsymbol{\Sigma}_{
        \boldsymbol{P}_{\mathcal{A}}
        \boldsymbol{P}_{\mathcal{A}}
    }^{+}
    \boldsymbol{\Sigma}_{
        \boldsymbol{Y}\boldsymbol{P}_{\mathcal{A}}
    }^{\mathsf{T}}$.
To accommodate the singular covariance matrices that may arise in classification, we define
$\widetilde{\boldsymbol{\Sigma}}_{\mathcal{A}}
    \triangleq
    \boldsymbol{\Sigma}_{
        \boldsymbol{E}_{\mathcal{A}}
        \boldsymbol{E}_{\mathcal{A}}
    } +
    \epsilon\mathbf{I}_C$, where $\epsilon=0$ when the covariance is nonsingular, and $\epsilon=10^{-8}$ is used otherwise for numerical regularisation.
Accordingly, the error vectors having Mahalanobis distance lower than $q \in \mathbb{R}^+$ are given by \cite{johnson2014applied}: $\mathcal{E}_{{\mathcal{A}}}
    =
    \{
    \boldsymbol{e}\in\mathbb{R}^{C}:
    \boldsymbol{e}^{\mathsf{T}}
    \widetilde{\boldsymbol{\Sigma}}_{\mathcal{A}}^{-1}
    \boldsymbol{e}
    \leq q^2
    \}$,
which defines a $C$-dimensional hyperellipsoid centered at $\mathbb{E}[\boldsymbol{E}_{\mathcal{A}}]=\boldsymbol{0}$, with volume $V_{\mathcal{A}}(q)
    =[{\pi^{C/2}q^C}/
         {\Gamma(C/2+1)}]\cdot[{\det} ( \widetilde{\boldsymbol{\Sigma}}_{\mathcal{A}}
        )]^{1/2}$. 
        
We define the \emph{conditional evidence gain} of completing sensor
$s_i$ as the logarithmic reduction of this volume:
\begin{equation}
    \widetilde{I}(\boldsymbol{Y};\widehat{\boldsymbol{Y}}_i|\boldsymbol{P}_{\mathcal{A}})
    \triangleq
    \frac{1}{2}
    \log_2\{
    {
        {\det}(
            \widetilde{\boldsymbol{\Sigma}}_{\mathcal{A}}
        )
    }/{
        {\det}(
            \widetilde{\boldsymbol{\Sigma}}_{\mathcal{A}\cup\{s_i\}}
        )
    }\}.
    \label{eq:conditional_evidence_gain}
\end{equation}
Unlike a fixed sensor-relevance score, $\widetilde{I}(\cdot)$
accounts for redundancy and complementarity: the value of sensor
$s_i$ changes with the evidence already contained in
$\mathcal{A}$.
When $\epsilon=0$ and $\boldsymbol{Y}$ and the sensor prediction
errors are jointly Gaussian, the conditional evidence gain
coincides with the conditional mutual information:
$\widetilde{I}(\boldsymbol{Y};\widehat{\boldsymbol{Y}}_i|\boldsymbol{P}_{\mathcal{A}})
    {=}
    I(\boldsymbol{Y};\widehat{\boldsymbol{Y}}_i|\boldsymbol{P}_{\mathcal{A}})$.
This follows from the entropy chain rule applied to $h(\boldsymbol{Y}\mid\boldsymbol{P}_{\mathcal{A}})$ and the fact that the conditional covariance is the Schur complement of $\boldsymbol{\Sigma}_{\boldsymbol{P}_{\mathcal{A}}\boldsymbol{P}_{\mathcal{A}}}$. We omit the proof for brevity.

The evidence gain in
\eqref{eq:conditional_evidence_gain} quantifies the benefit of completing sensor $s_i$'s data object but does not account for the
wireless resources required to acquire it. To combine these
aspects, consider the allocation of one \gls{rb} to sensor $s_i$
during slot $k$. Given its current \gls{se}
$\xi_{i,k}$, the fraction
of its remaining data ${b}_{i,k}$ served by that \gls{rb} is $\alpha_{i,k} \triangleq
    \min\{
        1,{\xi_{i,k}BT_s}/{{b}_{i,k}} \}$.
    
Accordingly, we define the \gls{mira} priority score as
\begin{equation}
    \Theta_{i,k}
    \triangleq
    \alpha_{i,k}
    \widetilde{I}(\boldsymbol{Y};\widehat{\boldsymbol{Y}}_i|\boldsymbol{P}_{\mathcal{A}_k}).
    \label{eq:mira_score}
\end{equation}
This score favours sensors that offer high conditional evidence
gain while requiring relatively little radio service to complete.
Since $\widetilde{I}(\cdot)$ measures the value of the
complete observation, \eqref{eq:mira_score} approximates its
one-\gls{rb} contribution by scaling it with the fraction of the
remaining data to be transmitted (assuming that task value accumulates proportionally with the received data fraction).
Prediction performance is nevertheless evaluated using the actual complete or imputed observations available when the \gls{twi} closes at $T$.

\vspace{+0.045cm}
\noindent
{\bf MIRA.} At the beginning of slot $k$, \gls{mira} initialises
$\widetilde{M}=M$. While $\widetilde{M}>0$ and at least one sensor in
$\mathcal{D}_k$ has positive \gls{se}, the
policy selects
\vspace{-0.35cm}
\begin{equation}
    s_{i}
    =
    \arg\hspace{-0,46cm}\max_{
        s_j\in\mathcal{D}_k:\,
        \xi_{j,k}>0
    }
    \Theta_{j,k}.
    \label{eq:mira_selection}
\end{equation}
The number of \glspl{rb} required to complete its buffer
within the slot is given by
$m_{i,k} \triangleq
\lceil {b}_{i,k}/(\xi_{i,k}BT_s) \rceil$. Then the policy allocates
$M_{i,k} = \min\{m_{i,k}, \widetilde{M}\}$.
If $m_{i,k}\leq\widetilde{M}$, sensor
$s_{i}$ is completed within the slot and the working sets
are updated as
$\mathcal{D}_k
    \leftarrow
    \mathcal{D}_k\setminus\{s_i\}$
    and
    ${\mathcal{A}_k}
    \leftarrow
    {\mathcal{A}_k}\cup\{s_i\}$.
The conditional gains of the remaining sensors are then updated
given the enlarged evidence set. Otherwise, all remaining
\glspl{rb} are assigned to $s_{i}$. In either case, $\widetilde{M}
    \leftarrow
    \widetilde{M}-M_{i,k}$, ${b}_{i,k}\leftarrow{b}_{i,k}-\xi_{i,k}M_{i,k}BT_s$,
and the procedure continues until all \glspl{rb} have been
allocated or no pending sensor can transmit. Repeating this
procedure at every slot allows \gls{mira} to adapt jointly to the
evolving wireless state and to the conditional value of the
evidence already acquired.

\vspace{+0.045cm}
\noindent
{\bf Benchmark Policies.} We compare \gls{mira} to two benchmark variants that follow the
same within-slot allocation procedure but use different sensor-priority
scores.

The first is \emph{\gls{mia}}, that we propose as an ablation variant of
\gls{mira}, and replaces \eqref{eq:mira_selection} with
$s_{i_{\mathrm{MIA}}} {=} \arg\max_{s_j\in\mathcal{D}_k}
\widetilde{I}(\boldsymbol{Y};\widehat{\boldsymbol{Y}}_j|\boldsymbol{P}_{\mathcal{A}_k})$.
\gls{mia} accounts only for the conditional value of each observation but
disregards the channel state information. The second is \emph{Best-CQI}, a maximum sum-rate policy~\cite{li2013ofdma} which replaces \eqref{eq:mira_selection}
with
$s_{i_{\mathrm{CQI}}} {=} \arg\max_{s_j\in\mathcal{D}_k} \xi_{j,k}$.
This prioritises transmission efficiency but does not consider the inferential value of the transmitted
observations.

\noindent
{\bf Complexity.}
At each allocation step, MIRA compares the priority scores
in~\eqref{eq:mira_score} of at most $S$ pending sensors. Since
each step either completes one sensor or assigns all remaining
\glspl{rb} to an incomplete sensor, a TTI contains at most
$\min\{M,S\}$ allocation steps, leading
to $\mathcal{O}(S\min\{M,S\})$ as selection complexity per TTI. When a sensor is completed,
the candidates' evidence gains must be updated. Each gain in \eqref{eq:conditional_evidence_gain} requires a pseudoinverse of dimension at
most $SC$, with complexity $\mathcal{O}((SC)^3)$. Hence if
$\mathcal{A}_k$ expands at every allocation step, the worst-case
complexity is
$\mathcal{O}(S\min\{M,S\}(SC)^3)$ per TTI.
\vspace{-0.1cm}
\section{Performance Evaluation}
\label{sec:numerical}
\begin{table}[!t]
    \centering
    \caption{Common wireless-system parameters}
    \label{tab:system_parameters}
    \renewcommand{\arraystretch}{1.08}
    \setlength{\tabcolsep}{2pt}
    \scriptsize
    \begin{tabular}{@{}l c @{\hspace{5pt}} l c@{}}
        \hline
        \textbf{Parameter} & \textbf{Value}
        & \textbf{Parameter} & \textbf{Value} \\
        \hline
        Total BW, $MB$
        & $1.08$ MHz
        & \gls{rb} BW, $B$
        & $180$ kHz \\

        Subcarrier spacing
        & $15$ kHz
        & Integration time, $T_{I,i}$
        & $\mathcal{U}[10,100]$ ms \\

        Slot duration, $T_s$
        & $1$ ms
        & Initial buffer, $b_i^\star$
        & $\mathcal{U}[1,100]$ kB \\

        Number of \glspl{rb}, $M$
        & $6$
        & Average \gls{snr}
        & $5$ dB \\
        \hline
    \end{tabular}\vspace{-0.3cm}
\end{table}
We evaluate \gls{mira} across three inference settings: synthetic
multimodal classification, human activity
classification, and vehicle-trajectory regression. In every
setting, \gls{mira} is compared to \gls{mia} and Best-CQI using the
same sensor-readiness and channel realisations. Results are reported as a function of the evidence-acquisition deadline $T$.

We consider two average-\gls{snr} configurations: \emph{uniform}, with $\gamma_i{=}5$ dB for every sensor, and \emph{rank-based}, in which sensors are ordered by their marginal evidence gain $\widetilde{I}(\boldsymbol{Y};\widehat{\boldsymbol{Y}}_j|\emptyset)$ (from least informative $s_1$ to most informative $s_S$) and assigned $\gamma_i=S-2i+1$ {dB},
a deliberately adverse setting in which the most informative sensors experience the poorest average channels. At the end of the \gls{twi}, missing elements of partially received observations are replaced with their training-set means and fused according to \eqref{eq:lmmse_fusion}. Besides task performance, we report the average percentage of data (in bits, normalised by $\sum_i b_i^\star$) received before $T$. Unless stated otherwise, wireless parameters follow Table~\ref{tab:system_parameters}: sensor integration times and initial buffer sizes are drawn independently per Monte Carlo realisation from the listed distributions, Rayleigh fading gains vary independently across sensors and slots, \gls{se} values follow the mapping of~\cite{markova2024performance}, and results are averaged over $1{,}000$ realisations.

\begin{figure}[t]
    \centering

    \includegraphics{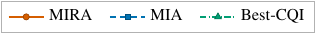}\\[1pt]

    \begin{subfigure}{\linewidth}
        \centering
        \includegraphics{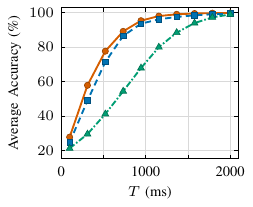}\hfil
        \includegraphics{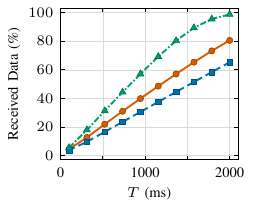}
        \vspace{-0.2cm}
        \caption{Uniform-based average \gls{snr}}
        \label{fig:acc_synth_uniform}
    \end{subfigure}

   \vspace{0.3cm}

    \begin{subfigure}{\linewidth}
        \centering
        \includegraphics{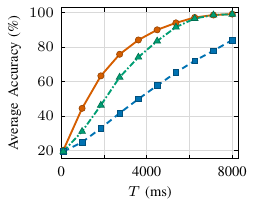}\hfil
        \includegraphics{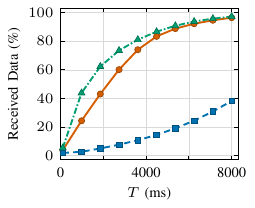}
        \vspace{-0.2cm}
        \caption{Rank-based average \gls{snr}}
        \label{fig:acc_synth_snr}
    \end{subfigure}

    \caption{{\small Synthetic classification under the two considered \gls{snr} scenarios. Left: prediction accuracy; Right: percentage of sensor data received by the evidence-acquisition deadline $T$.}}
    \label{fig:acc_synth}
    \vspace{-4mm}
\end{figure}

\vspace{+0.045cm}
\noindent
{\bf Synthetic Multimodal Classification.}
We consider a synthetic classification task with $C{=}6$ equiprobable classes and $F{=}25$ features partitioned among $S{=}10$ sensors, each observing $N_i{\in}\{2,3\}$ features. For class $c$, sensor $s_i$'s observation is drawn as $\boldsymbol{X}_i|c {\sim} \mathcal{N}(\boldsymbol{\mu}_{i,c},\mathbf{I}_{N_i})$ with $\boldsymbol{\mu}_{i,c} {\sim} \mathcal{N}(\boldsymbol{0},\lambda_i^2\mathbf{I}_{N_i})$, where $\lambda_i$ is linearly spaced over $[0.5,2.5]$ across sensors, with larger values emulating greater class separation and thus higher task relevance. A Naive Bayes classifier is trained independently per sensor modality, its posterior-probability vector forming $\widehat{\boldsymbol{Y}}_i$. The means and covariances required by the fusion rule are estimated from out-of-fold predictions via five-fold cross-validation. Under the uniform-\gls{snr} configuration ($\gamma_i{=}5$\,dB $\forall i$), Fig.~\ref{fig:acc_synth_uniform} shows accuracy and received-data percentage versus the acquisition horizon $T$. \gls{mira} achieves the highest accuracy, with relative improvements of up to $95.11\%$ over Best-CQI and $15.03\%$ over MIA. The advantage is largest for short-to-moderate $T$, when radio resources most restrict the evidence that can be collected, and the policies converge as $T$ grows since most observations eventually become receivable regardless of order. Best-CQI receives the largest data volume but not the highest accuracy; \gls{mia} receives up to $18.75\%$ less data than \gls{mira} and $35.00\%$ less than Best-CQI while nearly matching \gls{mira}'s accuracy, offering a favourable accuracy-throughput trade-off and potentially lower power consumption. These results underscore that delivered data volume alone is not a reliable measure of the inferential value of the resulting evidence. {From a temporal-requirement perspective, if we consider a target accuracy of $80\%$, Best-CQI achieves it at $T\approx1190$ ms, whereas \gls{mira} and \gls{mia} achieve it at $T\approx500$ ms, thus providing a reduction of $\sim58\%$.}

Under the rank-based \gls{snr} setup ($\gamma_i=S-2i+1$ dB), Fig.~\ref{fig:acc_synth_snr} shows that \gls{mira} still provides the highest accuracy, with relative improvements of up to $35.40\%$ over Best-CQI and $70.10\%$ over MIA. \gls{mia} prioritises highly informative (but costly to acquire) sensors, resulting in very low received data volume and, consequently, degraded accuracy. In contrast, Best-CQI favours efficiently transmitted but weakly relevant observations; \gls{mira} instead constructs more useful evidence by combining both factors.

\begin{figure}[t]
    \centering
    \includegraphics{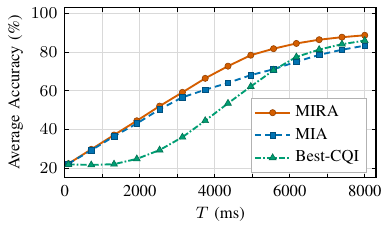}
    \vspace{-0.2cm}
    \caption{{\small Human activity classification accuracy under rank-based average \gls{snr} versus the evidence-acquisition horizon $T$.}\vspace{-0.1cm}}
    \label{fig:acc_har_snr}
\end{figure}

\vspace{+0.045cm}
\noindent
{\bf Real-World Human Activity Recognition.}
We consider the Human Activity Recognition dataset in~\cite{humanactivity}, with $F{=}60$ smartphone accelerometer features for $C{=}5$ activities randomly partitioned among $S=10$ sensors with $N_i{=}6$ features each. Each sensor-specific predictor is a feed-forward neural network with one hidden layer of $15$ ReLU units and a softmax output, trained for at most $100$ iterations. To emulate heterogeneous sensing quality, features are corrupted by additive white Gaussian noise (AWGN) with standard deviation $\sigma_i$ linearly spaced in $[0.2, 3.0]$ (in linear scale) across sensors. Under the rank-based \gls{snr} configuration, Fig.~\ref{fig:acc_har_snr} shows \gls{mira} achieving the highest accuracy throughout the considered horizons, with relative improvements of up to $66.70\%$ over Best-CQI and $17.15\%$ over \gls{mia}. The latter matches \gls{mira} for short and moderate horizons; Best-CQI approaches \gls{mia} only for longer horizons, when enough time allows acquiring more informative observations from sensors with slower transmission.

\begin{figure}[t]
    \centering
    \includegraphics{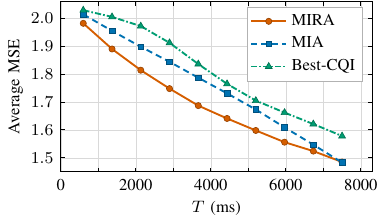}
    \vspace{-0.2cm}
    \caption{{\small Vehicle-trajectory prediction \gls{mse} under rank-based average \gls{snr} versus the evidence-acquisition horizon $T$. \vspace{-0.4cm} }}
    \label{fig:mse_reg}
\end{figure}

\vspace{+0.045cm}
\noindent
{\bf Vehicle-Trajectory Regression.}
To evaluate \gls{mira} beyond classification, we use the real-world NGSIM US-101 dataset~\cite{ngsim}, which provides nine features for each vehicle. The target is the vehicle position (horizontal and vertical) one second ahead, $\boldsymbol{Y} = [X_h(t+1), X_v(t+1)]^{\mathsf{T}}$. The $F=9$ features are treated as observations from $S=9$ sensor sources (one feature each). As in the human activity experiment, heterogeneous sensing quality is emulated by adding AWGN with $\sigma_i$ spaced in $[0.2,3.0]$. For every sensor and target component, we train a feed-forward regression model with one hidden layer of $15$ ReLU units and a linear output for at most $100$ iterations. Predictors and targets are standardised using training-set statistics, so the reported \gls{mse} is in standardised units. Under the rank-based \gls{snr} configuration, Fig.~\ref{fig:mse_reg} shows \gls{mira} achieving the lowest \gls{mse} throughout, with reductions of up to $9.09\%$ relative to Best-CQI and $5.52\%$ relative to MIA. Best-CQI performs worst consistently, since channel quality alone does not identify the observations most relevant to trajectory prediction; the gap between MIA and \gls{mira} narrows for longer horizons, as prioritising informative sensors with poor channels becomes less costly. 

These results show that the proposed allocation principle extends to both classification and regression tasks.

\vspace{-0.1cm}
\section{Conclusions}
\label{sec:conclusions}

This paper formulated multimodal uplink scheduling within a prescribed \gls{twi} as a wireless evidence-acquisition problem, in which radio-resource decisions progressively construct the evidence available for a common prediction. For \gls{lmmse} fusion, we derived an information-based metric from the reduction in residual error volume, which coincides with conditional mutual information under joint Gaussianity, and developed \gls{mira}, a low-complexity policy combining this gain with channel state information and remaining data volume. Across synthetic classification, human activity recognition, and vehicle-trajectory regression, \gls{mira} outperforms relevance-only and best-channel scheduling. Against Best-CQI, \gls{mira} achieves up to $95\%$ higher classification accuracy and $9\%$ lower regression \gls{mse}, while requiring up to $58\%$ less time to attain an $80\%$ target accuracy. Gains are more pronounced when modality relevance and channel quality are misaligned, confirming that maximising data need not maximise inference. Future work will explore nonlinear, learned, or dynamically adapted fusion mechanisms that better capture cross-modal dependencies, together with lower-complexity approximations of the evidence gain metric to enable real-time operation on constrained edge nodes.
\vspace{-0.1cm}
\bibliography{IEEEabrv,references}
\bibliographystyle{IEEEtran}

\end{document}